\documentclass{article}
\begin{document}
\title{Dark energy and dissipation}
\author{Diego Pav\'{o}n\\
Departamento de F\'{\i}sica, UAB,\\
08193 Bellaterra (Barcelona), Spain\\
Luis P. Chimento $\,$ \& $\,$ Alejandro S. Jakubi\\
Departamento de F\'{\i}sica. Universidad de Buenos Aires,\\
1428 Buenos Aires, Argentina}

\date{\today}
\maketitle

\begin{abstract}
Most of the models leading to a current state of cosmic accelerated
expansion fail to address the coincidence problem, i.e., that the
dark energy density and the energy density of the matter fluid are of
the same order precisely today. We show that a way to drive late acceleration
and simultaneously solve the aforesaid problem is assuming the
matter fluid dissipative \cite{enlarged}.%
\end{abstract}

\section{General setting}
The Friedmann equation plus the conservation equations for matter
and  dark energy in a Friedmann--Robertson--Walker universe
dominated by  these two components (non--interacting with one 
another), in terms of the density parameters, read
\\
\begin{eqnarray}
1&=&\Omega_{m}  +  \Omega_{\phi} + \Omega_{k} \, , \\
\dot{\Omega} & = & (3\gamma -2) H (\Omega -1) \Omega \, ,\\
\dot{\Omega}_{\phi} & = & [2+(3\gamma -2) \Omega -
3 \gamma_{\phi}]\Omega_{\phi} H \, ,
\label{eq:ekg} 
\end{eqnarray}
\\
where $\Omega \equiv \Omega_{m} + \Omega_{\phi}$, and $\gamma$ stands for
the overall adiabatic index 
$\gamma = (\gamma_{m} \Omega_{m} + \gamma_{\phi} \Omega_{\phi})/\Omega$,
with $\gamma_{m,\, \phi} \equiv 1 + (p_{m,\, \phi}/\rho_{m,\, \phi})$,
and such that    $1 \leq \gamma_{m} \leq 2$ and $0 \leq \gamma_{\phi} < 1 $ 
(it should be noted that in general $\gamma_{m}$ and $\gamma_{\phi}$ may 
vary with time).

It is not our aim to propose a new dark energy model but rather to show
that if one wishes to have late cosmic acceleration and simultaneosly
solve the coincidence problem using general relativity, then the matter 
fluid component must be dissipative -in the sense explained below-
irrespective of the potential of the scalar field \cite{enlarged}.
  
{}From the above equations it is immediately seen that for $\Omega = 1$
Eq.  (3) implies that $\dot{\Omega}_{\phi} > 0$. Consequently, at
large times $\Omega_{\phi} \rightarrow 1$ and $\Omega_{m} \rightarrow
0$, i.e., the accelerated expansion ($q <0 \Longrightarrow
\gamma_{\phi} < 2/3$) and the coincidence problem cannot be solved
simultaneously within this approach. Moreover, for the solution
$\Omega = 1$ to be stable the overall adiabatic index must comply
with the upper bound $\gamma < 2/3$ which is uncomfortably low.

A way out is to assume the matter fluid dissipative, i.e., endowed
with a nonequilibrium scalar pressure $\pi$. This is always negative
for expanding universes as required by the second law of thermodynamics 
\cite{murphy}, \cite{djd}, \cite{la}. Further, the occurrence of $\pi$ is 
rather natural since (exception made of superfluids) all matter fluids 
found in Nature are dissipative -see e.g. \cite{batchelor}, \cite{ll}. 
Thereby Eqs. (2) and (3) generalize to
\\
\begin{eqnarray}
\dot{\Omega} & = &  \left[3 \left(\gamma+ 
\frac{\pi}{\rho}\right) -2\right] H (\Omega -1) \Omega\, ,  \\
\dot{\Omega}_{\phi} & = & \left\{2+\left[3\left(\gamma + 
\frac{\pi}{\rho}\right) -2\right] \Omega -
3 \gamma_{\phi}\right\} H \Omega_{\phi}.
\label{eq:generalize} 
\end{eqnarray}
\\
We now may have $\dot{\Omega}_{\phi} < 0$ as well as $\Omega_{m}
\rightarrow \Omega_{m0} \neq 0$ and $\Omega_{\phi } \rightarrow
\Omega_{\phi 0} \neq 0$ for late time so long as the stationary
condition
\\
\begin{equation}
\gamma_{m} + \frac{\pi}{\rho_{m}} = \gamma_{\phi} = - 2 \frac{\dot{H}}
{3H^{2}}
\label{eq:attractor}
\end{equation}
\\
is satisfied. Besides, the constraint $\gamma < 2/3$ is replaced by
$\gamma + (\pi/\rho) < 2/3$, which is somewhat easier to fulfill.

\section{Stability analysis}
For spatially flat FRW universes the asymptotic stability of the
stationary solution $\Omega_{m0}$ and $\Omega_{\phi 0}$ can be 
studied from equation (\ref{eq:generalize}). By slightly perturbing 
$\Omega_{\phi}$ it follows that the solution is stable (and therefore 
an attractor) provided the quantity 
$\gamma_{m} + \frac{\pi}{\rho_{m}}- \gamma_{\phi}< 0 $ and tends to
zero as $t \rightarrow \infty$. This coincides with the stationary
condition (\ref{eq:attractor}).

For $\Omega \neq 1$ (i.e., when $k \neq 0 $), it is expedient to introduce
the ansatz $\epsilon = \epsilon_{0} + \delta$ in Eqs (4) \& (5), where  
$\epsilon_{0} \equiv (\Omega_{m}/\Omega_{\phi})_{0} \sim {\cal O}(1)$
and $\mid \delta \mid \ll \epsilon_{0}$. One finds that
\\
\begin{equation}
\dot{\delta} = - \frac{3}{\Omega_{\phi}}\left(\frac{2}{3} - \gamma_{\phi}
\right) \Omega_{k} H (\epsilon_{0} + \delta).
\label{delta}
\end{equation}
\\
As a consequence, the stationary solution will be stable for open FRW
universes ($\Omega_{k} > 0$). For closed FRW universes one has to go
beyond the linear analysis.

A realization of these ideas can be found in \cite{enlarged}. There it is 
seen that the space parameter is ample enough that no fine tuning is required 
to have late acceleration together with the fact that both density parameters 
tend to constant values compatible with observation.

Moreover we have shown that, for a wide class of dissipative dark energy
models, attractor solutions are themselves attracted towards a common
asymptotic behavior. This ``superattractor'' regime provides a model of the
recent universe that also exhibits an excellent fit to supernovae luminosity
observations and no age conflict \cite{qsa}.

\section{Concluding remarks}
Although dissipative pressures have been repeteadly invoked when
building cosmological models -see e.g. \cite{murphy}, \cite{caderni},
\cite{dw}, \cite{antf}, \cite{moscow}- one may ask about its origin in 
the present context. On the one hand if the  CDM particles are really 
self--interacting (as it has been recently argued \cite{spergel}), 
then they are bound to arise. (CDM self--interaction with mean free 
path between about $1$ kpc and $1$ Mpc seems to be key to explain the
structure of the halo of galaxies).  On the other hand, it is well
known that particle production somehow plays the role of a
dissipative pressure \cite{jdb}. The latter is linked to the rate
$\Gamma$ of particle production by $\pi = - \gamma_{m} \rho_{m}
\Gamma/(3H)$. Therefore, even a very small rate can produce a
significant dissipative pressure
\cite{winfried}.

According to the standard picture of cosmic structure formation a
sufficiently long matter--dominated era must have taken place during
which the observed structure grew from the density fluctuations
measured by CMB anisotropy experiments -see e.g. \cite{padmanabhan}.
As a consequence the transition to the currently observed accelerated
regime should have been quite recent in the evolution of the
universe.  This seems to imply that dissipative effects in dark
matter were small until density inhomogeneities became large. If so,
it is suggestive to think that the size of dissipative processes, as
measured by the ratio $\pi/\rho_{m}$, increased with dark matter
density and became large precisely because of the development of
inhomogeneities \cite{dominik}. The observed smoothness of halo
structure and the correlation between high energy cosmic ray
production and dark matter clustering might provide support 
for this relationship \cite{dick}.

Before closing, we would like to emphasize that our analysis does not
apply to scenarios in which the dark energy component interacts with
the matter component -see e.g. \cite{luca}, \cite{wdl}. This is easy to
understand as the interaction somehow mimics a dissipative pressure.

\section*{Acknowledgments}
This work has been partially supported by the
Spanish Ministry of Science and Technology under grant BFM  
2000-C-03-01 and 2000-1322, and the University of Buenos 
Aires under Project X223.


\begin{thebibliography}{99}
\bibitem{enlarged} Chimento, L.P., Jakubi, A.S., Pav\'{o}n, D., 2000,
Phys. Rev. D 62, 063508
\bibitem{murphy} Murphy, G.L., 1973, Phys. Rev. D 8, 4231
\bibitem{djd} Pav\'{o}n, D., Bafaluy, J., Jou, D., 1991, Class. Quantum 
Grav. 8, 347
\bibitem{la} Chimento, L.P., Jakubi, A.S., 1993, Class. Quantum 
Grav. 10, 2047
\bibitem{batchelor} Batchelor, G.K., 1967, {\it An introduction to
fluid dynamics}. Cambridge University Press, Cambridge
\bibitem{ll} Landau, L., Lifshitz, E., 1971 {\it M\'{e}canique des fluids}.
\'{E}ditions MIR, Moscou
\bibitem{qsa} Chimento L. P., Jakubi A. S. and Zuccal\'a N.A., 2001,
Phys. Rev. D 63, 103508
\bibitem{caderni} Caderni, N., Fabbri, R., 1979, Phys. Rev. D 20, 1254
\bibitem{dw} Pav\'{o}n, D., Zimdahl, W., 1993, Phys. Lett. A 179, 261
\bibitem{antf} Zimdahl, W., Schwarz, D.J., Balakin, A.B., Pav\'{o}n, D.,
Phys. Rev. D 64, 063501
\bibitem{moscow} Zimdahl, W., Schwarz, D.J., Balakin, A.B., Pav\'{o}n, D.,
2002, Gravitation and Cosmology 8, 158
\bibitem{spergel} Spergel, D.N. and Steinhardt, P.J., 2000, Phys.
Rev. Lett. 84, 3760
\bibitem{jdb} Barrow, J.D., 1988, Nucl. Phys. B 310, 743
\bibitem{winfried} Zimdahl, W., 2000, Phys. Rev. D 61, 083511
\bibitem{padmanabhan} Padmanabhan, T., 1993, {\it Structure formation in
the universe}, Cambridge University Press, Cambridge 
\bibitem{dominik} Schwarz, D., 2003, ``Accelerated expansion without
dark energy", in this volume, {\sf astro-ph/0209584}
\bibitem{dick} Dick, R., Blasi, P., Kolb, E.W., 2002, ``Ultrahigh energy
cosmic rays from dark matter annihilation", {\sf astro-ph/0205158}
\bibitem{luca} Amendola, L., 2000, Phys. Rev D 62, 043511
\bibitem{wdl} Zimdahl, W., Pav\'{o}n, D., Chimento, L.P., 2001
Phys. Lett. B 521, 133
\end{thebibliography}
\end{document}